\documentclass[journal]{IEEEtran}
\usepackage{amsmath,amsfonts,graphicx,tabularx,color}
\usepackage{amssymb,textcomp,mathtools, amsthm}
\usepackage{bm,upgreek,algorithm,hyperref}
\usepackage{multirow,booktabs,hhline,array}
\usepackage{cite,url,makecell,setspace,listings}
\RequirePackage{doi}

\def\thline{\noalign{\hrule height 1.0pt}}

\definecolor{codegreen}{rgb}{0,0.6,0}
\definecolor{codegray}{rgb}{0.5,0.5,0.5}
\definecolor{codepurple}{rgb}{0.58,0,0.82}
\definecolor{backcolour}{rgb}{0.95,0.95,0.92}

\lstdefinestyle{mystyle}{
    backgroundcolor=\color{backcolour},   
    commentstyle=\color{codegreen},
    keywordstyle=\color{magenta},
    numberstyle=\tiny\color{codegray},
    stringstyle=\color{codepurple},
    basicstyle=\ttfamily\scriptsize,
    breakatwhitespace=false,         
    breaklines=true,                 
    captionpos=b,                    
    keepspaces=true,                 
    numbers=left,                    
    numbersep=5pt,                  
    showspaces=false,                
    showstringspaces=false,
    showtabs=false,                  
    tabsize=2
}

\DeclareMathOperator*{\argmin}{argmin}

\renewcommand{\vec}[1]{\bm{\mathrm{#1}}}

\theoremstyle{definition}

\theoremstyle{plain}

\theoremstyle{remark}

\title{A Time-domain Real-valued Generalized Wiener Filter for Multi-channel Neural Separation Systems}

\author{Yi~Luo}

\begin{document}
\maketitle
\setlength{\abovedisplayskip}{2pt}
\setlength{\belowdisplayskip}{2pt}
\setlength{\abovedisplayshortskip}{2pt}
\setlength{\belowdisplayshortskip}{2pt}

\begin{abstract}
Frequency-domain beamformers have been successful in a wide range of multi-channel neural separation systems in the past years. However, the operations in conventional frequency-domain beamformers are typically independently-defined and complex-valued, which result in two drawbacks: the former does not fully utilize the advantage of end-to-end optimization, and the latter may introduce numerical instability during the training phase. Motivated by the recent success in end-to-end neural separation systems, in this paper we propose time-domain real-valued generalized Wiener filter (TD-GWF), a linear filter defined on a 2-D learnable real-valued signal transform. TD-GWF splits the transformed representation into groups and performs an minimum mean-square error (MMSE) estimation on all available channels on each of the groups. We show how TD-GWF can be connected to conventional filter-and-sum beamformers when certain signal transform and the number of groups are specified. Moreover, given the recent success in the sequential neural beamforming frameworks, we show how TD-GWF can be applied in such frameworks to perform iterative beamforming and separation to obtain an overall performance gain. Comprehensive experiment results show that TD-GWF performs consistently better than conventional frequency-domain beamformers in the sequential neural beamforming pipeline with various neural network architectures, microphone array scenarios, and task configurations.
\end{abstract}


\section{Introduction}
\label{sec:intro}
Recent studies on neural beamformers have significantly advanced the state-of-the-art of multi-channel speech enhancement \cite{yoshioka2015ntt, heymann2015blstm, qian2018deep, wang2018all}, speech separation \cite{yin2018multi, ochiai2020beam, wang2021sequential, zhang2021adl, ni2021wpd++}, and automatic speech recognition (ASR) systems \cite{xiao2016beamforming, xiao2016deep, ochiai2017unified, zhang2017speech, heymann2018performance, drude2018integrating, chang2019mimo}. A neural beamformer typically first applies a neural network to extract the target sources in the noisy observations, and then uses a beamformer module to perform spatial filtering. Despite a few studies that explored the effect of time-domain beamformers \cite{qian2018deep, luo2019fasnet}, frequency-domain beamformers such as the multi-channel Wiener filter (MCWF), the minimum-variance distortionless response (MVDR) beamformer, and the generalized eigenvalue (GEV) beamformer are the common choices since both the microphone array and target source characteristics can be estimated in the frequency domain in a much easier way \cite{gannot2017consolidated}.

Existing frequency-domain neural beamformers often estimate time-frequency (T-F) masks for the pre-separation stage \cite{heymann2015blstm}. However, prior works on time-domain single-channel speech separation have discussed the potential drawbacks for conventional T-F masking methods in single-channel speech separation task \cite{luo2018tasnet}. Similarly, while conventional frequency-domain beamformers were successful in a wide range of systems and tasks, there are still two main limitations within: the \textit{end-to-end optimization ability} and the \textit{complex-valued operations}. First, the operations in conventional beamformers are typically defined by a set of optimization problems, and such operations are independent from the signals and cannot be jointly optimized with the entire end-to-end separation pipeline. Second, with more and more recent works started to apply neural networks on complex-valued spectrograms, how to properly handle the the complex-valued features in the nonlinear transforms becomes an important problem. Although a common way is to concatenate the real and imaginary parts into a larger feature to jointly model them \cite{williamson2015complex, tan2019learning, wang2020complex, zhang2021adl}, there are also methods that use different modules for real and imaginary parts and mimic the behavior of complex-valued operations \cite{lee2017fully, choi2018phase, hu2020dccrn}. Moreover, when the frequency-domain beamformers are placed within a network and the gradient of the rest of the network is passed through the beamformer operations during the backpropagation phase, numerical instability in such complex-valued operations might affect the training and introduce additional difficulties in the optimization of the entire system \cite{chakrabarty2015numerical}.

Time-domain beamformers have been investigated for general array processing tasks \cite{dougherty2004advanced, jaeckel2006strengths}. However, due to the well-defined problem formulation and the simplicity and efficiency in the implementation, frequency-domain beamformers are the mainstream in multi-channel speech processing tasks \cite{heymann2015blstm}. Moreover, as T-F masking was one of the most important single-channel speech enhancement and separation techniques in the past decades \cite{deliang2006casa}, frequency-domain beamformers are naturally suitable for frequency-domain speech enhancement and separation systems. With the recent success of time-domain neural source separation systems \cite{stoller2018wave, venkataramani2018end, luo2019fasnet, luo2019conv, lluis2019end, zhang2020furcanext, tzinis2020sudo, zeghidour2020wavesplit, luo2020dual, nachmani2020voice, subakan2020attention}, it is natural to revisit the formulation and application of time-domain beamformers in the framework of end-to-end source separation. The DeepBeam system \cite{qian2018deep} has already shown that directly applying a conventional time-domain multi-channel Wiener filter to a set of outputs generated by a time-domain speech enhancement model, and the filter-and-sum network (FaSNet) \cite{luo2019fasnet} has shown that directly estimating the filter coefficient of a simple time-domain filter-and-sum beamformer is applicable. However, both systems did not modify the formulation where the beamforming process was defined as a standard filter-and-sum operation.

In this paper, we propose the time-domain real-valued generalized Wiener filter (TD-GWF) as an alternative to frequency-domain conventional beamformers for end-to-end multi-channel neural separation systems. Unlike conventional time-domain beamformers whose filter coefficients are defined as 1-D filters, TD-GWF calculates the filter coefficients on a \textit{learnable} 2-D feature similar to the single-channel time-domain neural separation systems. The signal transform utilized to generate the 2-D feature can either be pre-defined or fully learnable, and certain choices of the signal transform connect TD-GWF to conventional time-domain or frequency-domain filter-and-sum beamformers. The filter coefficients, which are now in the form of a 2-D matrix, are defined as the solution to an minimum mean-square error (MMSE) estimation on the learnable 2-D features of the multi-channel observations and an estimated target source. Moreover, we consider the multi-channel separation task in the \textit{sequential beamforming pipeline} \cite{wang2021sequential, chen2021beam, wang2021multi}, which contains a pre-separation module, a beamforming module, and a post-separation module. The pre-separation module first estimates the target source of a selected reference microphone, and then the beamforming module calculates the beamformed target source based on the estimation. The post-separation module takes the outputs from the pre-separation and beamforming modules as auxiliary inputs and performs separation again to obtain a refined output. The beamforming-refinement process can be repeated for multiple iterations to build a sequential pipeline. In this framework formulation, TD-GWF introduces a \textit{group-splitting} operation which not only decreases the computational complexity but also improves the overall separation performance. Comprehensive experiment results show that replacing the conventional frequency-domain beamformers by TD-GWF in the sequential neural beamforming pipeline drastically improves the separation performance across various microphone array scenarios and task configurations.

The rest of the paper is organized as follows. Section~\ref{sec:GWF} briefly overviews the conventional frequency-domain beamformers and introduces the proposed TD-GWF and its application in the sequential beamforming pipeline. Section~\ref{sec:config} provides the dataset and experiment configurations. Section~\ref{sec:result} presents the experiment results. Section~\ref{sec:conclusion} concludes the paper.

\section{Time-domain Real-valued Generalized Wiener Filter}
\label{sec:GWF}
\subsection{Recap of Conventional Frequency-domain Neural Beamformers}
\label{sec:freq-bf}

We start with a quick recap of a formulation of the conventional frequency-domain neural beamformers. Given $M$ channels of $L$-sample noisy observations $\{\vec{s}_m\}_{m=1}^M, \vec{s}_m \in \mathbb{R}^{1\times L}$, a neural network is first applied to either a selected channel (e.g. a reference channel) or all the channels to estimate the source-of-interest (SOI) $\hat{\vec{x}} \in \mathbb{R}^{Q\times L}$, where $Q \in \{1, M\}$ denotes the index of the outputs of the neural network. Most of the prior works apply standard frequency-domain filter-and-sum beamformers such as MWF, MVDR and GEV, where the signals are first transformed to frequency domain via short-time Fourier transform (STFT):
\begin{align}
\begin{split}
    \vec{S}_m &= \text{STFT}(\vec{s}_m) \\
    \hat{\vec{Z}}_q &= \text{STFT}(\hat{\vec{x}}_q)
\end{split}
\end{align}
where $\vec{S}_m, \hat{\vec{Z}}_q \in \mathbb{C}^{F\times T}$ correspond to the complex-valued spectrogram of the $m$-th observation and $q$-th estimated SOI, respectively, and $F$ and $T$ represent the number of frequency bins and time steps, respectively. The linear beamforming filter at frequency $f$ is typically defined as an $M$-dimensional complex-valued vector $\vec{h}(f) \in \mathbb{C}^{M\times 1}$ applied to the spectrograms of the observations:
\begin{align}
    \bar{\vec{z}}(f, t) = \vec{h}(f)^\dagger\vec{S}(f, t)
\end{align}
where $\vec{S}(f, t) \in \mathbb{C}^{M\times 1}$ denotes the time-frequency bins at $f$-th frequency and $t$-th frame in the spectrogram of all the $M$ channels, $\dagger$ denotes the conjugate transpose, and $\bar{\vec{z}}(f, t) \in \mathbb{C}$ denotes the beamformed time-frequency bin for the SOI. The estimation of $\vec{h}(f)$ can be done by solving certain optimization problems designed for various purposes. For example, the conventional frequency-domain MCWF (FD-MCWF) can be defined as the MMSE solution between the beamformed output and the spectrogram of the SOI $\hat{\vec{z}}(f, t)$ estimated by the neural network:
\begin{align}
\begin{split}
    \vec{h}_{\text{MCWF}}(f) &= \argmin_{\vec{h}} \mathbb{E}_t\,[||\vec{h}(f)^\dagger\vec{S}(f, t) - \hat{\vec{z}}(f, t)||_2] \\
    &= \mathbb{E}_t[\vec{S}(f, t)\vec{S}(f, t)^\dagger]^{-1}\mathbb{E}_t[\vec{S}(f, t)\hat{\vec{z}}(f, t)^\dagger]
\label{eqn:MCWF}
\end{split}
\end{align}

\subsection{Time-domain Real-valued Generalized Wiener Filter}
\label{sec:TD-RV-GWF}

\begin{figure*}[!t]
	\small
	\centering
	\includegraphics[width=2\columnwidth]{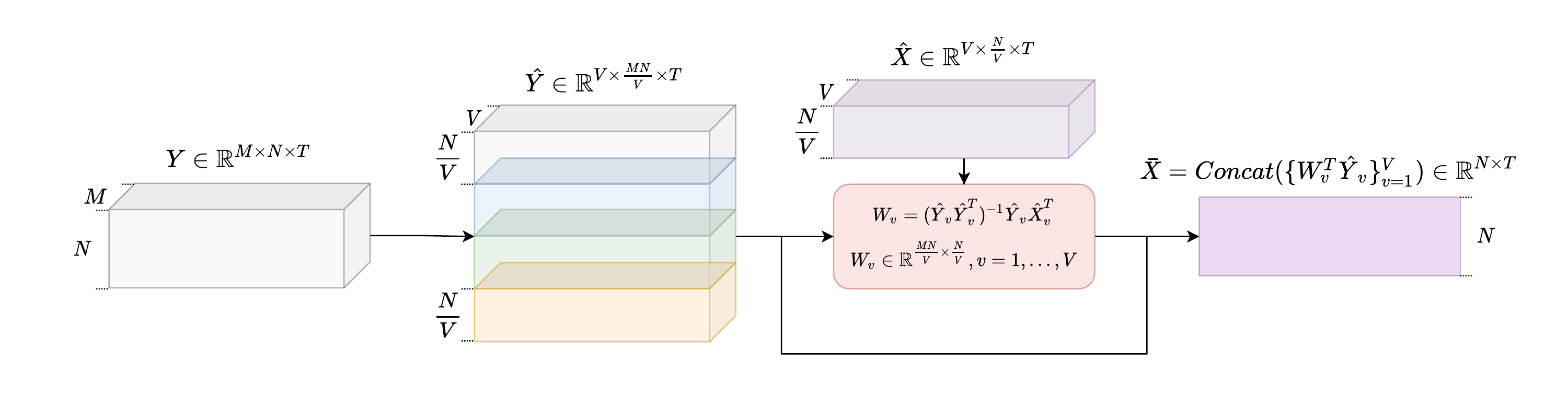}
	\caption{Flowchart of the proposed time-domain generalized Wiener filter (TD-GWF). The spectrograms of the $M$ observations $\vec{Y}$ and the estimated SOI $\hat{\vec{X}}$ are split into $V$ groups along the frequency axis, and then$\vec{Y}$ are concatenated along the group dimension to form another tensor $\hat{\vec{Y}}$. For each group $v$ in $\hat{\vec{Y}}$ and $\hat{\vec{X}}$, an MMSE estimation is performed between $\hat{\vec{Y}}_v$ and $\hat{\vec{X}}_v$ to calcualte the generalized Wiener filter coefficients $\vec{W}_v$. $\vec{W}_v$ is then applied to $\hat{\vec{Y}}_v$, and the outputs from all the $V$ groups are concatenated along the group axis to form the final estimation of the SOI $\bar{\vec{X}}$.}
	\label{fig:flowchart}
\end{figure*}

The proposed time-domain real-valued generalized Wiener filter (TD-GWF) is a linear filter defined on a \textit{learnable} real-valued signal transform beyond STFT. Inspired from the recent success in time-domain source separation systems, we replace the complex-valued Fourier transform in the derivation of conventional frequency-domain beamformers to a learnable real-valued linear transform which can be jointly optimized with the other parts of the separation system.

Figure~\ref{fig:flowchart} shows the procedure of the proposed TD-GWF. Similar to the use of STFT in conventional frequency-domain beamformers, we transform the 1-D waveform signals into 2-D features by applying a real-valued linear transform:
\begin{align}
\begin{split}
    \vec{Y}_{m,t} &= \vec{y}_{m,t}\vec{B} \\
    \hat{\vec{X}}_t &= \hat{\vec{x}}_t\vec{B}
\end{split}
\end{align}
where $\vec{y}_{m,t}, \hat{\vec{x}}_t \in \mathbb{R}^{1\times P}$ denote the $t$-th frame of the windowed waveform with $P$ sample points at the $m$-th observation or the estimated SOI, respectively, $\vec{B} \in \mathbb{R}^{P\times N}$ denotes the linear transformation matrix or the real-valued waveform encoder that can be either pre-defined or jointly optimized with the entire system, and $\vec{Y}_m, \hat{\vec{X}} \in \mathbb{R}^{N\times T}$ denote the $N$-dimension sequential features of the $m$-th observation or the estimated SOI, respectively. Note that this is identical to the learnable signal encoders in recent time-domain speech separation systems such as the time-domain audio separation network (TasNet) \cite{luo2019conv}.

The 2-D features $\vec{Y}_m$ are then split to $V$ non-overlapped groups of $\frac{N}{V}$-dimension sub-features, and the $M$ channels of sub-features in the same group are then concatenated to form $V$ groups of transformed features of shape $\hat{\vec{Y}} \in \mathbb{R}^{V\times \frac{MN}{V} \times T}$. The same group-splitting process is also applied to $\hat{\vec{X}}$ to transform it to shape $\mathbb{R}^{V\times \frac{N}{V} \times T}$. Each group in $\hat{\vec{Y}}$ and $\hat{\vec{X}}$, denoted by $\hat{\vec{Y}}_v \in \mathbb{R}^{\frac{MN}{V} \times T}$ and $\hat{\vec{X}}_v \in \mathbb{R}^{\frac{N}{V} \times T}$, respectively, are used to calculate a Wiener filter $\vec{W}_v \in \mathbb{R}^{\frac{MN}{V}\times \frac{N}{V}}$ via MMSE estimation:
\begin{align}
    \vec{W}_v = \argmin_{\vec{W}_v} ||\vec{W}_v^\top\hat{\vec{Y}}_v - \hat{\vec{X}}_v||_2, \, v = 1,\ldots,V
\label{eqn:TD-GWF}
\end{align}
Unlike conventional frequency-domain beamformers, the estimation of $\vec{W}_v$ only depends on real-valued matrices.

$\vec{W}_v$ is then applied to $\hat{\vec{Y}}_v$ to obtain the $v$-th group of the output:
\begin{align}
    \bar{\vec{X}}_v = \vec{W}_v^\top\hat{\vec{Y}}_v
\end{align}

The final output $\bar{\vec{X}} \in \mathbb{R}^{N\times T}$ is obtained by concatenating the $V$ groups of outputs $\{\bar{\vec{X}}_v\}_{v=1}^V$ across the feature dimension:
\begin{align}
    \bar{\vec{X}} = \text{Concat}(\{\bar{X}_v\}_{v=1}^V)
\end{align}

A learnable signal decoder $\vec{D} \in \mathbb{R}^{N\times P}$ is then applied to $\bar{\vec{X}}$ to transform the 2-D representation back to the 1-D waveform $\bar{\vec{x}} \in \mathbb{R}^{1\times L}$:
\begin{align}
    \bar{\vec{x}} = \text{OLA}(\vec{D}^\top\bar{\vec{X}})
\end{align}
where $\text{OLA}(\cdot)$ represents the overlap-add operation on the windows.

\subsection{Choice of the Signal Transform}
\label{sec:transform}

The are multiple choices for the design of the real-valued learnable signal transform matrices $\vec{B}$ and $\vec{D}$, and here we provide three possible options.

\subsubsection{Identity Transform}
\label{sec:identity-transform}

The simplest signal transform is identity mapping, i.e., to directly use the waveforms of the mixtures and the SOI to calculate the TD-GWF coefficients. In this case, we have $\vec{B}=\vec{D}=\vec{I}$ and $N=P$, and the signal transform operation is equivalent to a simple windowing operation. More specifically, when $V=1$ and $\vec{W}$ is defined as a square Toeplitz matrix, the TD-GWF can be connected to a P-point time-domain filter-and-sum beamformer \cite{benesty2018multichannel}.

\subsubsection{Learnable Orthonormal Transform}

Another option for designing the learnable signal transform while both maintaining the \textit{perfect signal reconstruction} ability (i.e., $\vec{B}\vec{D}=\vec{I}$) and mimicking the frequency-independent behavior of STFT is to use a pair of real-valued orthonormal matrices for $\vec{B}$ and $\vec{D}$. To allow such orthonormal matrices to be learnable, we utilize the Householder transform (HHT) \cite{householder1958unitary} with a set of learnable real-valued vectors $\vec{v}_k \in \mathbb{R}^{1\times P},\,k=1,\ldots, K$:
\begin{align}
\begin{split}
    \hat{\vec{v}}_k &= \frac{\vec{v}_k}{||\vec{v}_k||_2} \\
    \vec{V}_k &= \vec{I} - 2\hat{\vec{v}}_k\hat{\vec{v}}_k^\top
\end{split}
\end{align}
The signal transform matrices $\vec{B}, \vec{D} \in \mathbb{R}^{P\times P}$ are then defined as:
\begin{align}
\label{eqn:HHT}
\begin{split}
    \vec{B} &= \vec{V}_1\vec{V}_2\cdots\vec{V}_K \\
    \vec{D} &= \vec{B}^{-1} = \vec{B}^\top
\end{split}
\end{align}
During the training phase, $\{\vec{v}_k\}_{k=1}^K$ are jointly optimized with the rest of the system. More specifically, when $\vec{B}$ is set to the discrete Fourier transform (DFT) matrix (in which case the real-valued constraint is no longer valid) and the number of groups $V$ is set to the window length $P$, $\vec{W}_v \in \mathbb{C}^{M\times 1}$ can be connected to the conventional complex-valued FD-MCWF.

\subsubsection{Learnable Unconstrained Transform}

The third option is to adopt a similar configuration as the single-channel time-domain source separation systems, which is to use unconstrained matrices for the signal transform. In this case, $\vec{B}$ and $\vec{D}$ are randomly initialized and jointly optimized with the rest of the system.

\subsection{TD-GWF in End-to-end Sequential Beamforming Pipeline}
\label{sec:GWF-seq}

\begin{figure*}[!t]
	\small
	\centering
	\includegraphics[width=1.4\columnwidth]{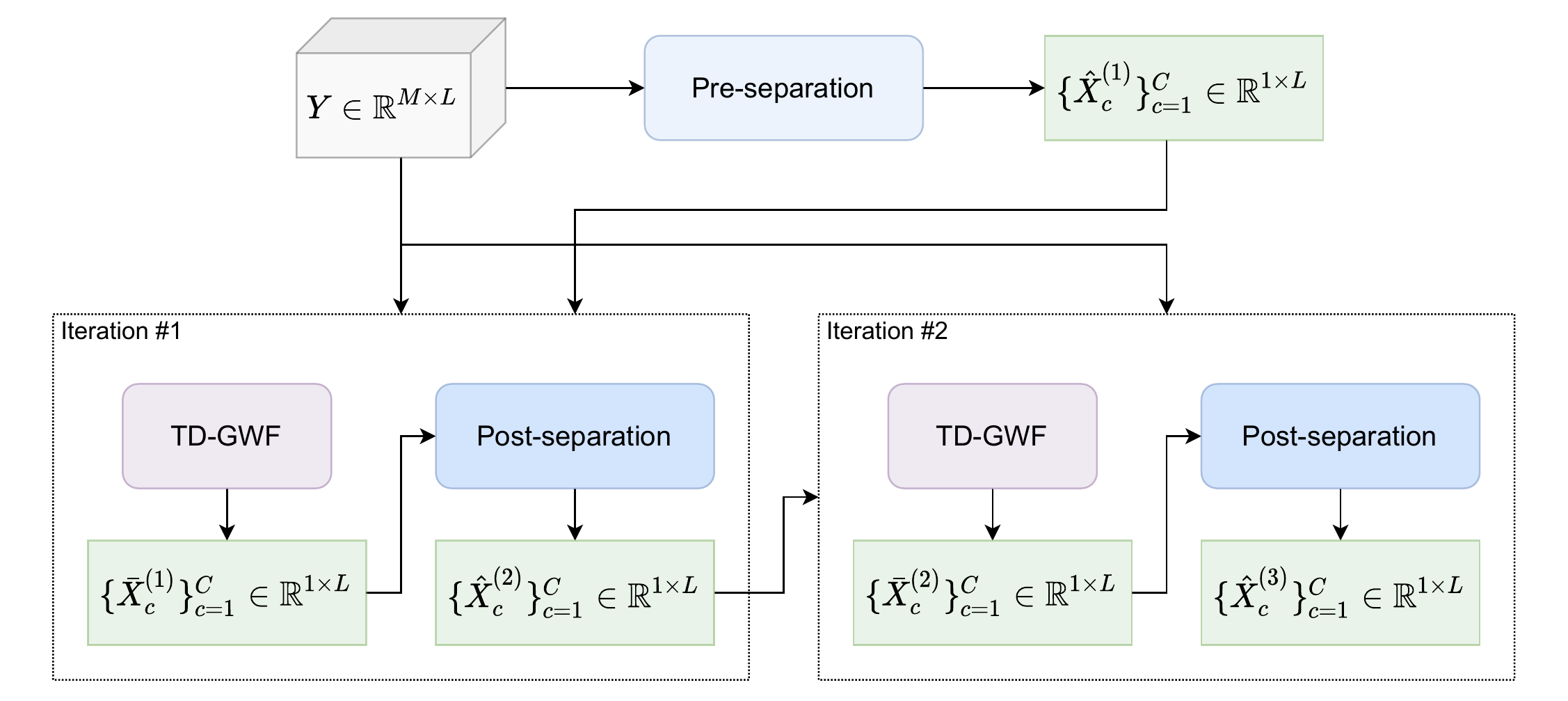}
	\caption{Flowchart for the sequential neural beamforming pipeline. The pre-separation module first estimates the target sources in a selected reference microphone, and then TD-GWF is applied to calculate the filtered estimated targets. The post-separation module takes the outputs from the pre-separation and TD-GWF module as auxiliary inputs and perform separation again. The beamforming-refinement process can be repeated for multiple iterations.}
	\label{fig:iterative}
\end{figure*}

Although TD-GWF can be directly applied to obtain the estimated SOI $\bar{\vec{x}}$, we find it more powerful when inserted to \textit{sequential beamforming pipelines}. A general design of sequential neural beamforming pipelines contains a \textit{pre-separation} module, a \textit{beamformer}, and a \textit{post-enhancement} module \cite{wang2021sequential, chen2021beam}. The pre-separation module first performs separation on the noisy observations to obtain a coarse estimation of the SOIs, and then the beamformer uses those estimations to calculate the beamformer coefficients. The beamformed outputs, typically together with the coarse estimations from the pre-separation module, are then sent to a post-enhancement module for further refinements. The output from the post-enhancement module can be further sent to the beamforming module again for another round of processing, and such beamforming-refinement procedure can be repeated to form a sequential pipeline. Current pipelines have investigated the use of FD-MCWF and FD-MVDR beamformers and reported significant performance improvements compared to separator-only or single-stage neural beamforming baselines \cite{ochiai2020beam, wang2021sequential}.

Figure~\ref{fig:iterative} shows the pipeline of TD-GWF-based sequential neural beamformer. We follow the general design of \cite{wang2021sequential} and we replace the FD-MCWF by the proposed TD-GWF. The pre-separation module first takes the observations $\vec{Y}$ as input and generates the $C$ estimated SOIs $\{\hat{\vec{x}}_c^{(1)}\}_{c=1}^C \in \mathbb{R}^{1\times L}$ at a selected reference microphone. We then calculate the TD-GWF output $\{\bar{\vec{x}}_c^{(1)}\}_{c=1}^C$ based on the procedure described in Section~\ref{sec:TD-RV-GWF}. The post separation module is another single-channel separation network which takes $\vec{Y}$, $\{\hat{\vec{x}}_c^{(1)}\}_{c=1}^C$ and $\{\bar{\vec{x}}_c^{(1)}\}_{c=1}^C$ as input and generates the refined SOIs $\{\hat{\vec{x}}_c^{(2)}\}_{c=1}^C$. For an iterative beamforming-refinement process, the refined output at stage $j\geq 2$, i.e., $\{\hat{\vec{x}}_c^{(j)}\}_{c=1}^C$, is sent to the TD-GWF module to generate $\{\bar{\vec{x}}_c^{(j)}\}_{c=1}^C$, and $\vec{Y}$, $\{\hat{\vec{x}}_c^{(j)}\}_{c=1}^C$ and $\{\bar{\vec{x}}_c^{(j)}\}_{c=1}^C$ are sent to the post-separation module again to generate $\{\hat{\vec{x}}_c^{(j+1)}\}_{c=1}^C$. The post-separation module is shared by all iterations.

The training of the system can be done by applying the training objective function to the outputs from all separation modules:
\begin{align}
        \mathcal{L}_{\text{obj}} = \frac{1}{K} \sum_{k=1}^K D_\Pi(\{\hat{\vec{x}}_c^{(k)}\}_{c=1}^C, \{\vec{x}_c\}_{c=1}^C)
    \end{align}
where $C$ denotes the total number of SOIs, $K$ denotes the number of beamforming-refinement iterations, $D(\cdot)$ is a selected loss function, and $D_\Pi(\cdot)$ denotes permutation invariant training (PIT) \cite{yu2017permutation}. During inference phase, either $\{\hat{\vec{x}}_c^{(K)}\}_{c=1}^C$ or $\{\bar{\vec{x}}_c^{(K)}\}_{c=1}^C$ can be used as the final output.

\subsection{Rules of Thumb in Implementation}
\label{sec:rules}

Conventional frequency-domain beamformers may encounter difficulties such as numerical stability during the training of end-to-end multi-channel neural separation systems \cite{zhang2021end}. Although TD-GWF does not involve any complex-valued operations thus bypasses several potential issues in frequency-domain beamformers, we still find a few rules of thumb in the implementation that allows the system to be faster and more robust during training and achieve better performance during inference.

\subsubsection{Solving the MMSE problem in TD-GWF}

The MMSE problem in equation~\ref{eqn:TD-GWF} has a closed-form solution:
\begin{align}
    \vec{W}_v = \text{pinv}(\hat{\vec{Y}}_v)\hat{\vec{X}}_v^\top = (\hat{\vec{Y}}_v\hat{\vec{Y}}_v^\top)^{-1}\hat{\vec{Y}}_v\hat{\vec{X}}_v^\top
\end{align}
where $\text{pinv}(\cdot)$ denotes the pseudo-inverse operator. While we can directly calculate the pseudo-inverse of $\hat{\vec{Y}}_v$ or the matrix inverse of $\hat{\vec{Y}}_v\hat{\vec{Y}}_v^\top$, we empirically find that methods that provides faster and more numerically stable least-square solutions, e.g., the $\textit{solve}$ function\footnote{\url{https://pytorch.org/docs/1.10/generated/torch.linalg.solve.html}} and the $\textit{lstsq}$ function\footnote{\url{https://pytorch.org/docs/1.10/generated/torch.linalg.lstsq.html}}, can also stabilize the training of the entire system. This also matches the previous observations in frequency-domain beamformers \cite{zhang2021end}. A Pytorch-style code snippet is provided as follows for calculating $\vec{W}_v$:
\begin{lstlisting}[language=Python]
def calc_W(Y_v, X_v):
    # Y_v: shape (B, M*N, T)
    # X_v: shape (B, N, T)
    
    Y_cov = Y_v.bmm(Y_v.transpose(1,2))  # (B, M*N, M*N)
    YX = Y_v.bmm(X_v.transpose(1,2))  # (B, M*N, N)
    W_v = torch.linalg.solve(Y_cov, YX)  # (B, M*N, N)
    return W_v
\end{lstlisting}

\subsubsection{Stop-gradient Operation in Sequential Beamforming Pipeline}

We train the sequential beamforming pipeline by applying the training objective function on the outputs of all iterations. This is based on the assumption that improving the quality of the outputs from the previous iteration will help the performance of both the TD-GWF and the post-separation modules in the current iteration. Previous studies on iterative source separation pipelines have shown that when the outputs at previous iterations are used as auxiliary inputs for the current iteration, it is better to detach the gradient of the previous outputs so that the gradients are constrained within the iteration \cite{luo2021empirical}. Since the post-separation module is shared across iterations, the intuition for such stop-gradient operation is that the optimization of the post-separation module in the current iteration should not affect the optimization of it in prior iterations. The same applies to the TD-GWF module (i.e., the signal transform matrices $\vec{B}$ and $\vec{D}$), as they are also shared across iterations. We thus detach the gradients of $\{\hat{\vec{x}}_c^{(j)}\}_{c=1}^C$ when they are sent to the TD-GWF module in the next iteration, and in this configuration the signal transform $\vec{B}$ and $\vec{D}$ for TD-GWF is jointly optimized with the post-separation module within each iteration. Note that other sequential beamforming systems also apply such constraint during training to achieve a good performance \cite{wang2021sequential}.

\section{Experiment configurations}
\label{sec:config}
\subsection{Dataset}

We use the same dataset proposed in \cite{luo2020end} for evaluating the effectiveness of the proposed TD-GWF. The simulated dataset contains 20000, 5000 and 3000 4-second long utterances sampled at 16 kHz sample rate for training, validation and test sets, respectively, and both ad-hoc array and fixed circular array configurations are utilized. For each utterance, two speech signals and one noise signal are randomly selected from the 100-hour Librispeech subset \cite{panayotov2015librispeech} and the 100 Nonspeech Corpus \cite{web100nonspeech}, respectively. The overlap ratio between the two speakers is uniformly sampled between 0\% and 100\%, and the two speech signals are shifted accordingly and rescaled to a random relative SNR between 0 and 5 dB. The relative SNR between the power of the sum of the two clean speech signals and the noise is randomly sampled between 10 and 20 dB. The transformed signals are then convolved with the room impulse responses (RIRs) simulated by the image method \cite{allen1979image} using the gpuRIR toolbox \cite{diaz2020gpurir} for all microphones. The length and width of all the rooms are randomly sampled between 3 and 10 meters, and the height is randomly sampled between 2.5 and 4 meters. The reverberation time (T60) is randomly sampled between 0.1 and 0.5 seconds. After convolution, the echoic signals are summed to create the mixture for each microphone. For the ad-hoc array configuration, the number of microphones varies from 2 to 6. For the fixed circular array configuration, the number of microphones is set to 6 and the diameter of the array is set to 10~cm. The positions of the microphones as well as the sources are randomly placed inside the room with the constraint that they are at least 0.5~m away from the boarders. The data simulation scripts are publicly available online\footnote{\url{https://github.com/yluo42/TAC}}.

\begin{table*}[!ht]
	\scriptsize
	\centering
	\caption{Oracle performance of FD-MCWF, FD-PMWF, and the proposed GWF in both frequency-domain and time-domain. Oracle reverberant SOIs are used for the calculation of the beamformer coefficients.}
	\label{tab:oracle}
	\begin{tabular}{c|c|c|cc|cc}
		\thline
		\multirow{3}{*}{Method} & \multirow{3}{*}{Window} & \multirow{3}{*}{Group} & \multicolumn{2}{c}{SDR (dB)} & \multicolumn{2}{c}{SI-SDR (dB)} \\
		\cline{4-7}
		 & & & \multirow{2}{*}{Fixed} & Ad-hoc & \multirow{2}{*}{Fixed} & Ad-hoc \\
		 & & & & 2 / 4 / 6 mics & & 2 / 4 / 6 mics \\
		 \hline
		 Mixture & -- & -- & -0.4 & -0.3 / -0.4 / -0.4 & -0.5 & -0.4 / -0.5 / -0.4 \\
		 \hline
		 \multirow{5}{*}{FD-MCWF} & 32 ms & \multirow{5}{*}{--} & 3.0 & 1.6 / 1.2 / 1.5 & 0.2 & -2.0 / -2.0 / -1.4 \\
		  & 64 ms & & 5.2 & 2.8 / 3.6 / 4.8 & 3.5 & 0.2 / 1.6 / 3.2 \\
		  & 128 ms & & 8.3 & 3.9 / 6.3 / 8.8 & 7.4 & 2.2 / 5.3 / 8.1\\
		  & 256 ms & & 12.1 & 5.1 / 9.6 / 13.5 & 11.7 & 4.2 / 9.1 / 13.2 \\
		  & 512 ms & & 15.4 & 6.0 / 12.0 / 17.6 & 15.2 & 5.5 / 11.8 / 17.5 \\
		  \hline
		  FD-PMWF ($\beta=0$) & \multirow{4}{*}{512 ms} & \multirow{4}{*}{--} & 7.8 & 4.7 / 8.1 / 8.2 & 6.7 & 4.2 / 7.1 / 7.1 \\
		  FD-PMWF ($\beta=1$) & & & 7.9 & 5.6 / 8.1 / 8.1 & 6.8 & 4.8 / 7.2 / 7.0 \\
		  FD-PMWF ($\beta=5$) & & & 7.8 & 5.0 / 8.0 / 8.1 & 6.7 & 3.4 / 7.0 / 7.0 \\
		  FD-PMWF ($\beta=10$) & & & 7.7 & 4.3 / 7.8 / 8.0 & 6.5 & 2.3 / 6.7 / 6.9 \\
		  \hline
		  & \multirow{3}{*}{2 ms} & 1 & 7.2 & 3.8 / 5.3 / 6.5 & 6.1 & 2.9 / 4.2 / 5.4 \\
		  & & 2 & 5.3 & 2.8 / 3.8 / 4.6 & 4.2 & 2.0 / 2.9 / 3.7 \\
		  & & 4 & 3.4 & 1.6 / 2.3 / 2.9 & 2.5 & 1.0 / 1.6 / 2.2 \\
		  \cline{2-7}
		  & \multirow{3}{*}{4 ms} & 1 & 9.7 & 5.2 / 7.4 / 9.3 & 8.7 & 4.1 / 6.3 / 8.3 \\
		  & & 2 & 7.1 & 3.8 / 5.3 / 6.5 & 6.1 & 2.9 / 4.3 / 5.5 \\
		  TD-GWF & & 4 & 4.7 & 2.5 / 3.4 / 4.2 & 3.8 & 1.8 / 2.6 / 3.4 \\
		  \cline{2-7}
		  (Identity transform) & \multirow{3}{*}{8 ms} & 1 & 13.7 & 7.7 / 11.3 / 15.0 & 13.1 & 6.7 / 10.5 / 14.4 \\
		  & & 2 & 9.8 & 5.4 / 7.7 / 9.8 & 9.1 & 4.4 / 6.8 / 9.0 \\
		  & & 4 & 6.5 & 3.6 / 5.0 / 6.2 & 5.7 & 2.7 / 4.1 / 5.4 \\
		  \cline{2-7}
		  & \multirow{3}{*}{16 ms} & 1 & 30.8 & 13.9 / 21.4 / 42.1 & 30.8 & 13.4 / 21.4 / 42.1 \\
		  & & 2 & 16.6 & 8.7 / 14.0 / 20.0 & 16.4 & 8.0 / 13.6 / 19.7 \\
		  & & 4 & 10.2 & 5.4 / 8.1 / 10.6 & 9.7 & 4.6 / 7.5 / 10.2 \\
		\thline
	\end{tabular}
\end{table*}

\subsection{Model configurations}
\label{sec:model-config}

We select one single-channel model and three multi-channel models as the benchmark systems:
\begin{enumerate}
    \item DPRNN-TasNet \cite{luo2020dual}: DPRNN-TasNet follows the same system design as the standard TasNet model while use dual-path RNN (DPRNN) blocks for the separator. A dual-path RNN block contains an \textit{intra-chunk} RNN and an \textit{inter-chunk} RNN which iteratively process the sequential feature in local and global scales. Such dual-path architecture has also shown effective with other network architectures \cite{chen2020dual, nachmani2020voice, subakan2021attention}, and here we select the RNN-based architecture due to its simplicity.
    \item MC-TasNet \cite{gu2019end}: The multi-channel TasNet (MC-TasNet) system extends the single-channel TasNet system by either extracting various cross-channel features \cite{gu2019end, gu2020enhancing} or using extra waveform encoders. Here we select the \textit{parallel encoder} configuration where each input channel has its own waveform encoder, and the encoded features are concatenated to serve as the input to the separator. Since the features are concatenated, MC-TasNet is suitable for fixed geometry array scenario where the microphone indices are known in advance.
    \item FaSNet-TAC \cite{luo2020end}: The filter-and-sum network (FaSNet) with transform-average-concatenate (TAC) module is a multi-channel end-to-end separation system designed for microphone permutation and number invariant scenarios. FaSNet estimated time-domain filter-and-sum beamforming coefficient with a neural network, and the TAC module incorporates the cross-channel features in a permutation invariant way. The FaSNet-TAC system can be applied to both fixed geometry array and ad-hoc array scenarios.
    \item iFaSNet \cite{luo2020implicit}: The implicit filter-and-sum network (iFaSNet) is a variant to the FaSNet-TAC system which performs implicit filter-and-sum on the features generated by a learnable waveform encoder. Different cross-channel features and filtering process have also been proposed to replace the original setting in FaSNet-TAC system. The iFaSNet system is mainly proposed for ad-hoc microphone array scenario.
\end{enumerate}
We encourage the readers to refer to the corresponding literature for details about these architectures. DPRNN blocks are also selected for the separators in MC-TasNet, FaSNet-TAC, and iFaSNet. Each system contains a small (marked as ``-S'') and a large (marked as ``-L'') setting, where the MC-TasNet systems contains 3 and 6 DPRNN blocks in the small and large settings, respectively, and both FaSNet-TAC and iFaSNet contain 2 and 4 DPRNN blocks in the small and large settings, respectively.

For the sequential beamforming pipeline, all systems above can be used for the pre-separation module. We always use the small setting for the pre-separation module. For the post-separation module, we always use the single-channel DPRNN-TasNet with the small setting, and the only difference is that $\vec{Y}$, $\{\hat{\vec{x}}_c^{(j)}\}_{c=1}^C$ and $\{\bar{\vec{x}}_c^{(j)}\}_{c=1}^C$ described in Section~\ref{sec:GWF-seq} are all encoded by the waveform encoder, and the encoded sequential features are concatenated to serve as the input to the DPRNN blocks. 

The three options for the signal transform matrices $\vec{B}$ and $\vec{D}$ in TD-GWF, described in Section~\ref{sec:transform}, are compared in the sequential beamforming pipeline. For the learnable orthonormal transform option, we empirically find that the number of Householder transforms $K$ in equation~\ref{eqn:HHT} do not lead to a significant difference in the final performance, and we set $K$ to 2 in all experiments. For the learnable unconstrained transform option, we set $\vec{B}$ and $\vec{D}$ to be square matrices for a fair comparison with the other two options. The window size $P$ for $\vec{B}$ and $\vec{D}$ ranges from 32 (2~ms) to 4096 (512~ms), and the number of groups $V$ ranges from 1 to $P$. Note that the waveform encoder and decoder in the separation modules are always different from the signal transform matrices in the TD-GWF module. The hop size of the signal transform is set to 25\% of the window size for all time-domain and frequency-domain beamformers. We always use a Hanning window for frequency-domain beamformers, and do not use any analysis window for TD-GWF.

\subsection{Training and Evaluation}

All models are trained for 100 epochs with the Adam optimizer \cite{kingma2014adam} with an initial learning rate of 0.001. Signal-to-noise ratio (SNR) is used as the training objective $D(\cdot)$. The learning rate is decayed by 0.98 for every two epochs. Gradient clipping by a maximum gradient norm of 5 is applied. We report the signal-to-distortion ratio (SDR) \cite{vincent2006performance}, the scale-invariant signal-to-distortion ratio (SI-SDR) \cite{Roux2019SDR}, and the wideband perceptual evaluation of speech quality (PESQ) \cite{rix2001perceptual} for signal quality evaluation.

\section{Results and analysis}
\label{sec:result}
\begin{table*}[!ht]
	\scriptsize
	\centering
	\caption{Comparison of different models on the simulated 6-mic circular array. Identity signal transform is used for TD-GWF-based models. SI-SDR is reported on decibel scale.}
	\label{tab:fixed}
	\begin{tabular}{c|c|c|c|c|cccc|cccc|c}
		\thline
		\multirow{2}{*}{Model} & \multirow{2}{*}{\# of param.} & System & \multirow{2}{*}{\# of iter.} & Inference & \multicolumn{4}{c|}{Speaker angle} & \multicolumn{4}{c|}{Overlap ratio} & \multirow{2}{*}{Average} \\
		\cline{6-13}
		& & output & & speed & $<$15\textdegree & 15-45\textdegree & 45-90\textdegree & $>$90\textdegree & $<$25\% & 25-50\% & 50-75\% & $>$75\% & \\
		\hline
		Mixture & -- & -- & -- & -- & -0.5 & -0.4 & -0.4 & -0.4 & -0.4 & -0.4 & -0.5 & -0.4 & -0.4 \\
		\hline
		DPRNN-TasNet-S & 1.3M & \multirow{2}{*}{--} & \multirow{2}{*}{--} & 19.2~ms & 7.8 & 8.1 & 8.5 & 8.7 & 13.2 & 9.4 & 6.7 & 3.9 & 8.3 \\
		DPRNN-TasNet-L & 2.6M & & & 36.5~ms & 8.2 & 8.5 & 8.8 & 9.0 & 13.4 & 9.7 & 7.0 & 4.4 & 8.6 \\
		\hline
		& 1.3M & \multirow{2}{*}{FD-MCWF} & 1 & 84.0~ms & -4.1 & -3.0 & -2.1 & -1.8 & -1.2 & -2.4 & -3.3 & -4.1 & -2.8 \\
		FD-MCWF-TasNet & 2.6M & & 2 & 130.6~ms & -3.7 & -2.5 & -1.4 & -1.1 & -0.9 & -2.0 & -2.5 & -3.4 & -2.2 \\
		\cline{2-14}
		(32 ms) & 1.3M & \multirow{2}{*}{Post-sep} & 1 & 103.1~ms & 8.9 & 9.3 & 9.9 & 10.2 & 14.3 & 10.8 & 8.0 & 5.2 & 9.6 \\
		& 2.6M & & 2 & 148.6~ms & 9.3 & 9.9 & 10.4 & 10.8 & 14.8 & 11.2 & 8.5 & 5.9 & 10.1 \\
		\hline
		& 1.3M & \multirow{2}{*}{FD-MCWF} & 1 & 94.1~ms & 2.2 & 2.8 & 3.2 & 3.3 & 6.8 & 3.9 & 1.6 & -0.9 & 2.9 \\
		FD-MCWF-TasNet & 2.6M & & 2 & 141.2~ms & 4.0 & 4.7 & 5.0 & 5.2 & 8.5 & 5.8 & 3.6 & 1.0 & 4.7 \\
		\cline{2-14}
		(512 ms) & 1.3M & \multirow{2}{*}{Post-sep} & 1 & 113.8~ms & 10.1 & 10.3 & 10.5 & 11.0 & 15.4 & 11.8 & 8.8 & 6.0 & 10.5 \\
		& 2.6M & & 2 & 160.5~ms & \textbf{11.4} & \textbf{11.9} & 12.0 & 12.5 & \textbf{16.5} & 13.2 & 10.6 & 7.4 & 11.9 \\
		\hline
		& 1.3M & \multirow{2}{*}{TD-GWF} & 1 & 54.3~ms & 3.3 & 4.2 & 5.5 & 6.0 & 5.6 & 5.3 & 4.5 & 3.6 & 4.7 \\
		TD-GWF-TasNet & 2.6M & & 2 & 101.5~ms & 3.8 & 5.0 & 6.5 & 7.2 & 5.9 & 6.0 & 5.6 & 5.0 & 5.6 \\
		\cline{2-14}
		(2 ms) & 1.3M & \multirow{2}{*}{Post-sep} & 1 & 71.4~ms & 9.8 & 10.9 & 12.0 & 12.8 & 15.4 & 12.5 & 10.1 & 7.4 & 11.3 \\
		& 2.6M & & 2 & 119.8~ms & 10.2 & 11.5 & 12.8 & 13.5 & 15.8 & 13.0 & 10.7 & \textbf{8.6} & 12.0 \\
		\hline
		& 1.3M & \multirow{2}{*}{TD-GWF} & 1 & 52.9~ms & 4.6 & 5.7 & 6.8 & 7.3 & 7.5 & 6.7 & 5.8 & 4.3 & 6.1 \\
		TD-GWF-TasNet & 2.6M & & 2 & 103.6~ms & 5.3 & 6.8 & 8.3 & 8.9 & 8.1 & 7.8 & 7.3 & 6.1 & 7.3 \\
		\cline{2-14}
		(4 ms) & 1.3M & \multirow{2}{*}{Post-sep} & 1 & 72.5~ms & 10.0 & 11.1 & 11.9 & 12.6 & 15.6 & 12.6 & 10.1 & 7.3 & 11.4 \\
		& 2.6M & & 2 & 122.8~ms & 10.7 & \textbf{11.9} & \textbf{12.9} & \textbf{13.6} & 16.3 & \textbf{13.3} & \textbf{11.0} & 8.5 & \textbf{12.3} \\
		\hline
		 & 1.3M & \multirow{2}{*}{TD-GWF} & 1 & 69.8~ms & 6.0 & 6.9 & 7.6 & 7.8 & 9.9 & 8.0 & 5.9 & 4.4 & 7.0 \\
		TD-GWF-TasNet & 2.6M & & 2 & 135.9~ms & 7.4 & 8.8 & 9.7 & 10.2 & 11.1 & 10.0 & 8.3 & 6.7 & 9.0 \\
		\cline{2-14}
		(8 ms) & 1.3M & \multirow{2}{*}{Post-sep} & 1 & 89.3~ms & 10.1 & 10.9 & 11.5 & 12.0 & 15.5 & 12.3 & 9.5 & 7.2 & 11.1 \\
		& 2.6M & & 2 & 156.7~ms & 10.8 & 11.8 & 12.4 & 13.1 & 16.0 & 13.2 & 10.7 & 8.2 & 12.1 \\
		\thline
	\end{tabular}
\end{table*}

\subsection{Comparison of Oracle Performance}

We start with the comparison of oracle performance of TD-GWF and conventional frequency-domain beamformers when the clean reverberant SOI is assumed available. Here we use the identity signal transform for GWF for the evaluation. Table~\ref{tab:oracle} shows the oracle performance of various beamformer configurations on both the fixed geometry and ad-hoc array scenarios. We select the conventional FD-MCWF described in Section~\ref{sec:freq-bf} and the frequency-domain parameterized multi-channel Wiener filter (FD-PMWF) \cite{souden2009optimal} as the frequency-domain beamformers. The FD-PMWF here is defined as:
\begin{align}
\begin{split}
    \vec{h}_{\text{PMWF}}(f) &= \argmin_{\vec{h}}\, \mathbb{E}_t\, [||\vec{h}(f)^\dagger\vec{Z}(f, t) - \vec{z}_{ref}(f, t)||_2]\\
    &\qquad \qquad \qquad+ \beta\mathbb{E}_t\, [||\vec{h}(f)^\dagger\vec{N}(f, t)||_2] \\
    &= \mathbb{E}_t[\vec{Z}(f, t)\vec{Z}(f, t)^\dagger + \beta \vec{N}(f, t)\vec{N}(f, t)^\dagger ]^{-1} \\
    &\quad \,\, \mathbb{E}_t[\vec{Z}(f, t)\vec{z}_{ref}(f, t)^\dagger]
\end{split}
\end{align}
where $\vec{Z}(f, t), \vec{N}(f, t) \in \mathbb{C}^{M\times 1}$ correspond to the $M$-channel SOI and interference at frequency $f$ and frame $t$, respectively. The term $\beta \in \mathbb{R}, \beta \geq 0$ controls the balance between interference reduction and distortion control.

We first notice that for the frequency-domain beamformers, a large window size is important to achieve a good performance under SDR and SI-SDR metrics. Although this observation looks contradictory to the window size used in various prior works on ASR \cite{heymann2017beamnet, drude2018integrating, chang2019mimo}, we find that there are a few recent literature also showed the importance of a large window size to achieve a satisfying performance on signal-level metrics such as SDR and SI-SDR \cite{ochiai2020beam, chen2021beam}. This indicates that source separation tasks might require a different window size compared to ASR tasks under different evaluation metrics. We then notice that the FD-MCWF performs better than FD-PMWF with various configurations of $\beta$, and even with $\beta=0$. Since FD-PMWF requires the estimation of SOIs at all channels and FD-MCWF only requires the estimation of SOIs at the reference channel, this observation indicates that the simpler configuration of only performing separation on the reference channel is able to achieve better signal quality after beamforming than performing separation on all channels. We next find that the proposed TD-GWF with a 2~ms window achieves comparable performance as FD-MCWF with a 32~ms window, and TD-GWF with a 8~ms window achieves comparable performance as FD-MCWF with a 512~ms window. Moreover, increasing the number of groups $V$ decreases the number of available coefficients in the TD-GWF filter, and the oracle performance drops as a consequence. The results show that when evaluated by signal quality metrics, TD-GWF can achieve a better oracle performance with a much smaller window size.

\subsection{Performance of Different System Configurations on Fixed Geometry Array}

We then compare the performance of different systems on the fixed geometry array configuration. Table~\ref{tab:fixed} provides the performance of the single-channel benchmark system and the sequential beamforming pipelines with different beamformers, system outputs, and configurations. All sequential beamforming pipelines use the single-channel DPRNN-TasNet-S model architecture for the pre-separation and post-separation modules for a fair comparison, and the model size of all sequential beamforming pipelines match that of the DPRNN-TasNet-L system. The final output of the sequential beamforming pipelines can be either the output from the beamformers or the output from the post-separation module, and the performance of the pipelines with 1 iteration using the output from the post-separation module can be directly compared to that of the DPRNN-TasNet-L system since the only difference is whether the beamforming output is used as an auxiliary input for the entire system. We set the number of groups $V$ to 1 for TD-GWF.

We can see that FD-MCWF-based models perform relatively bad with beamforming outputs, which matches the observations on the oracle performances. Moreover, the performance of the systems with 2 iterations is consistently better than those with 1 iteration, which shows the effectiveness of the sequential beamforming pipeline. On the other hand, the TD-GWF-based model with 2~ms window size can achieve better performance than FD-MCWF-based model with 512~ms window size when the beamforming output is selected, and the TD-GWF-based model with 4~ms window size can achieve better performance than FD-MCWF-based model with 512~ms window size when the post-separation output is selected. What we can learn from the results is that the output of the beamforming module, no matter which beamformer we select, is served as an additional feature to the post-separation module that explicitly captures cross-channel information at utterance level. Hence the main purpose for the beamformer module in such sequential beamforming pipelines is not to improve the signal quality of its output, but to serve as a cross-channel feature extractor to further improve the performance of a second-stage neural separator. From this perspective, TD-GWF can do a better job on cross-channel feature extraction compared to FD-MCWF.

We also provide the inference speed of different systems measured on a single 4-second long sentence with a Nvidia-T4 GPU. The speed is averaged over 3000 trials. We observe that compared to frequency-domain beamformers which require complex-valued matrix operations, TD-GWF can be faster across all choices of window sizes. However, we would also like to note that the actual speed of the systems may vary on different computational platforms.

\begin{table*}[!ht]
	\scriptsize
	\centering
	\caption{Effect of different choices of waveform encoder/decoder and group size for TD-GWF with 4 ms window size. The result on reverberant separation task is reported.}
	\label{tab:fixed-enc}
	\begin{tabular}{c|c|c|cccc|cccc|c}
		\thline
		Encoder/ & \multirow{2}{*}{\# of iter.} & \multirow{2}{*}{Group} & \multicolumn{4}{c|}{Speaker angle} & \multicolumn{4}{c|}{Overlap ratio} & \multirow{2}{*}{Average} \\
		\cline{4-11}
		Decoder & & & $<$15\textdegree & 15-45\textdegree & 45-90\textdegree & $>$90\textdegree & $<$25\% & 25-50\% & 50-75\% & $>$75\% & \\
		\hline
		\multirow{14}{*}{Identity} & \multirow{7}{*}{1} & 1 & 10.0 & 11.1 & 11.9 & 12.6 & 15.6 & 12.6 & 10.1 & 7.3 & 11.4 \\
		& & 2 & 9.6 & 10.8 & 11.8 & 12.6 & 15.5 & 12.2 & 9.8 & 7.2 & 11.2 \\
		& & 4 & 9.3 & 10.3 & 11.6 & 12.3 & 15.2 & 11.8 & 9.6 & 6.8 & 10.8 \\
		& & 8 & 8.9 & 10.1 & 11.3 & 12.3 & 14.9 & 11.6 & 9.3 & 6.7 & 10.6 \\
		& & 16 & 8.9 & 10.0 & 11.3 & 12.2 & 14.9 & 11.7 & 9.3 & 6.6 & 10.6 \\
		& & 32 & 8.7 & 9.7 & 11.1 & 12.1 & 14.9 & 11.4 & 9.0 & 6.4 & 10.4 \\
		& & 64 & 8.4 & 9.1 & 9.9 & 10.2 & 14.1 & 10.4 & 7.9 & 5.2 & 9.4 \\
		\cline{2-12}
		 & \multirow{7}{*}{2} & 1 & 10.7 & 11.9 & 12.9 & 13.6 & 16.3 & 13.3 & 11.0 & 8.5 & 12.3 \\
		& & 2 & 10.4 & 11.7 & 12.8 & 13.6 & 16.0 & 13.1 & 10.8 & 8.6 & 12.1 \\
		& & 4 & 10.2 & 11.5 & 12.7 & 13.6 & 15.9 & 12.9 & 10.6 & 8.6 & 12.0 \\
		& & 8 & 9.4 & 10.8 & 12.1 & 13.2 & 15.4 & 12.4 & 10.0 & 7.5 & 11.3 \\
		& & 16 & 9.1 & 10.5 & 11.8 & 12.9 & 15.2 & 11.9 & 9.6 & 7.5 & 11.0 \\
		& & 32 & 8.8 & 10.0 & 11.5 & 12.5 & 14.9 & 11.6 & 9.3 & 6.8 & 10.7 \\
		& & 64 & 8.4 & 9.2 & 10.0 & 10.2 & 14.1 & 10.5 & 7.9 & 5.3 & 9.4 \\
		\hline
		\multirow{14}{*}{LOT} & \multirow{7}{*}{1} & 1 & 10.2 & 11.2 & 12.0 & 12.6 & 15.6 & 12.7 & 10.2 & 7.5 & 11.5 \\
		& & 2 & 9.7 & 10.8 & 11.8 & 12.5 & 15.4 & 12.2 & 9.9 & 7.2 & 11.2 \\
		& & 4 & 9.4 & 10.4 & 11.6 & 12.5 & 15.3 & 12.0 & 9.6 & 7.0 & 11.0 \\
		& & 8 & 8.9 & 10.1 & 11.3 & 12.1 & 14.9 & 11.8 & 9.1 & 6.6 & 10.6 \\
		& & 16 & 8.9 & 10.1 & 11.4 & 12.2 & 14.9 & 11.7 & 9.3 & 6.6 & 10.6 \\
		& & 32 & 8.8 & 10.1 & 11.4 & 12.3 & 14.9 & 11.7 & 9.2 & 6.6 & 10.6 \\
		& & 64 & 8.7 & 9.8 & 10.6 & 10.8 & 14.4 & 10.9 & 8.4 & 6.2 & 10.0 \\
		\cline{2-12}
		 & \multirow{7}{*}{2} & 1 & \textbf{11.0} & \textbf{12.2} & 13.0 & 13.8 & \textbf{16.4} & \textbf{13.5} & \textbf{11.2} & \textbf{8.8} & \textbf{12.5} \\
		& & 2 & 10.6 & 11.9 & 12.9 & 13.7 & 16.3 & 13.3 & 11.0 & 8.4 & 12.3 \\
		& & 4 & 9.7 & 11.1 & 12.3 & 13.2 & 15.6 & 12.6 & 10.2 & 7.7 & 11.5 \\
		& & 8 & 9.5 & 10.7 & 12.5 & 12.9 & 15.4 & 12.2 & 9.8 & 7.7 & 11.3 \\
		& & 16 & 9.8 & 11.2 & 12.5 & 13.5 & 15.8 & 12.7 & 10.4 & 8.1 & 11.7 \\
		& & 32 & 9.3 & 10.5 & 12.0 & 12.8 & 15.3 & 12.1 & 9.7 & 7.4 & 11.1 \\
		& & 64 & 8.9 & 9.8 & 10.9 & 11.0 & 14.6 & 11.2 & 8.6 & 6.2 & 10.2 \\
		\hline
		\multirow{14}{*}{LUT} & \multirow{7}{*}{1} & 1 & 9.6 & 10.3 & 11.3 & 12.0 & 15.1 & 11.9 & 9.5 & 6.8 & 10.8 \\
		& & 2 & 10.0 & 11.0 & 11.9 & 12.6 & 15.7 & 12.5 & 10.0 & 7.3 & 11.4 \\
		& & 4 & 10.1 & \textbf{11.4} & 12.3 & 12.9 & \textbf{15.9} & 12.8 & 10.4 & 7.6 & 11.7 \\
		& & 8 & 10.0 & \textbf{11.4} & \textbf{12.5} & 12.9 & 15.8 & 12.8 & \textbf{10.5} & 7.6 & 11.7 \\
		& & 16 & 10.0 & \textbf{11.4} & 12.4 & 13.0 & \textbf{15.9} & 12.8 & 10.4 & \textbf{7.8} & 11.7 \\
		& & 32 & \textbf{10.2} & \textbf{11.4} & 12.4 & \textbf{13.2} & \textbf{15.9} & \textbf{12.9} & \textbf{10.5} & \textbf{7.8} & \textbf{11.8} \\
		& & 64 & 9.4 & 10.6 & 11.7 & 11.8 & 15.1 & 12.0 & 9.5 & 6.9 & 10.9 \\
		\cline{2-12}
		& \multirow{7}{*}{2} & 1 & 10.3 & 11.3 & 12.4 & 13.1 & 15.8 & 12.9 & 10.5 & 7.8 & 11.8 \\
		& & 2 & 10.6 & 12.0 & 12.8 & 13.4 & 16.2 & 13.2 & 10.9 & 8.4 & 12.2 \\
		& & 4 & 10.6 & 12.0 & 13.1 & 13.9 & 16.3 & 13.4 & 11.1 & 8.7 & 12.4 \\
		& & 8 & 10.6 & 12.1 & 13.1 & \textbf{14.0} & 16.3 & \textbf{13.5} & \textbf{11.2} & 8.7 & 12.4 \\
		& & 16 & 10.5 & 12.1 & 13.1 & \textbf{14.0} & 16.3 & 13.4 & \textbf{11.2} & \textbf{8.8} & 12.4 \\
		& & 32 & 10.6 & 12.1 & \textbf{13.2} & \textbf{14.0} & 16.3 & \textbf{13.5} & \textbf{11.2} & \textbf{8.8} & 12.4 \\
		& & 64 & 9.7 & 11.2 & 12.1 & 12.3 & 15.4 & 12.3 & 10.0 & 7.5 & 11.3 \\
		\thline
	\end{tabular}
\end{table*}

\subsection{Performance of Models with Different Signal Transforms, Window Sizes, and the Number of Groups}

Starting from now, we always select the output from the post-separation module as the final output of the sequential beamforming system. Table~\ref{tab:fixed-enc} shows the performance of the TD-GWF-based sequential beamforming pipeline with different signal transforms and number of groups $V$. The window size for the signal transform is set to 4~ms ($P=64$) as this is the best configuration in Table~\ref{tab:fixed}. The ``LOT'' and ``LUT'' in the table represent the learnable orthonormal transform and learnable unconstrained transform, respectively. The models with identity encoder/decoder and LOT encoder/decoder have comparable performance, and the ones with LOT encoder/decoder are slightly better in multiple configurations of $V$. We also notice that the performance of models with identity encoder/decoder and LOT encoder/decoder drops as $V$ increases, which also matches our previous observations on the oracle performance. However, we find that for the LUT encoder/decoder, increasing $V$ leads to a performance improvement until $V=P/2$. Note that the LUT encoder/decoder do not enforce perfect signal reconstruction by definition, and a large $V$ further harms the oracle performance of TD-GWF. One possible explanation for the improved performance is that when the TD-GWF module and the post-separation are jointly optimized, LUT encoder/decoder can further improve the cross-channel feature extraction ability compared to other signal transforms, and since the DPRNN-TasNet system estimates element-wise multiplicative masks on the 2-D features encoded by its waveform encoder, a large $V$ improves the modeling ability of the post-separation module on estimating the masks for each feature dimension. We also notice that all encoder/decoder choices achieve their worst performance with $V=P=64$. One possible explanation is that setting $V=P$ makes the TD-GWF similar to conventional frequency-domain beamformers where the filter coefficients are individually estimated at each feature dimension, which not only harms the oracle performance but also makes the joint optimization of the signal transform and the post-separation module harder.

\begin{table*}[!ht]
	\scriptsize
	\centering
	\caption{Effect of different choices of waveform encoder/decoder, window size, and group size for TD-GWF. The result on reverberant separation task is reported.}
	\label{tab:fixed-config}
	\begin{tabular}{c|c|c|c|cccc|cccc|c}
		\thline
		Encoder/ & \multirow{2}{*}{Window} & \multirow{2}{*}{\# of iter.} & \multirow{2}{*}{Group} & \multicolumn{4}{c|}{Speaker angle} & \multicolumn{4}{c|}{Overlap ratio} & \multirow{2}{*}{Average} \\
		\cline{5-12}
		Decoder & & & & $<$15\textdegree & 15-45\textdegree & 45-90\textdegree & $>$90\textdegree & $<$25\% & 25-50\% & 50-75\% & $>$75\% & \\
		\hline
		\multirow{7}{*}{Identity} & \multirow{3}{*}{8 ms} & \multirow{3}{*}{1} & 16 & 9.1 & 10.1 & 11.4 & 12.3 & 15.0 & 11.8 & 9.3 & 6.7 & 10.7 \\
		 & & & 32 & 8.8 & 9.9 & 11.4 & 12.4 & 14.8 & 11.6 & 9.1 & 7.0 & 10.6 \\
		 & & & 64 & 8.7 & 9.7 & 11.0 & 11.9 & 14.8 & 11.4 & 8.9 & 6.3 & 10.3 \\
		\cline{2-13}
		 & \multirow{4}{*}{16 ms} & \multirow{4}{*}{1} & 1 & 8.3 & 8.7 & 8.9 & 9.2 & 13.6 & 9.9 & 7.3 & 4.3 & 8.8 \\
		 & & & 32 & 9.0 & 10.1 & 11.3 & 12.1 & 15.0 & 11.6 & 9.0 & 6.8 & 10.6 \\
		 & & & 64 & 8.7 & 9.8 & 11.1 & 12.1 & 14.8 & 11.4 & 9.1 & 6.3 & 10.4 \\
		 & & & 128 & 8.7 & 9.7 & 11.0 & 11.9 & 14.8 & 11.3 & 8.8 & 6.3 & 10.3 \\
		\hline
		\multirow{15}{*}{LUT} & \multirow{4}{*}{8 ms} & \multirow{4}{*}{1} & 1 & 9.7 & 10.1 & 10.7 & 11.2 & 14.9 & 11.6 & 9.0 & 6.1 & 10.4 \\
		 & & & 16 & 10.6 & 11.6 & 12.4 & 13.1 & 16.1 & 13.1 & 10.7 & 7.9 & 11.9 \\
		 & & & 32 & 10.6 & 11.9 & 12.8 & 13.4 & 16.2 & 13.2 & 10.8 & 8.5 & 12.2 \\
		 & & & 64 & 10.6 & 12.0 & \textbf{12.9} & 13.5 & 16.1 & 13.3 & 10.9 & 8.6 & 12.2 \\
		 \cline{2-13}
		 & \multirow{4}{*}{16 ms} & \multirow{4}{*}{1} & 1 & 8.3 & 8.7 & 9.0 & 9.3 & 13.7 & 10.0 & 7.2 & 4.4 & 8.8 \\
		 & & & 32 & \textbf{11.2} & 12.1 & 12.8 & \textbf{13.5} & 16.5 & 13.5 & \textbf{11.1} & \textbf{8.6} & \textbf{12.4} \\
		 & & & 64 & 11.1 & \textbf{12.2} & 12.8 & 13.4 & 16.4 & 13.5 & 11.0 & 8.5 & 12.3 \\
		 & & & 128 & 11.1 & \textbf{12.2} & \textbf{12.9} & \textbf{13.5} & 16.5 & 13.6 & \textbf{11.1} & 8.3 & \textbf{12.4} \\
		 \cline{2-13}
		 & 32 ms & \multirow{5}{*}{1} & 256 & \textbf{11.2} & \textbf{12.2} & 12.8 & \textbf{13.5} & \textbf{16.7} & \textbf{13.7} & \textbf{11.1} & 8.2 & \textbf{12.4} \\
		 & 64 ms & & 512 & 11.1 & 11.9 & 12.5 & 13.0 & 16.4 & 13.4 & 11.0 & 7.7 & 12.1 \\
		 & 128 ms & & 1024 & 10.4 & 11.2 & 11.7 & 12.2 & 15.8 & 12.7 & 10.1 & 6.9 & 11.4 \\
		 & 256 ms & & 2048 & 7.9 & 8.5 & 9.3 & 9.6 & 13.7 & 10.1 & 7.2 & 4.3 & 8.8 \\
		 & 512 ms & & 4096 & 7.3 & 7.5 & 7.7 & 8.0 & 12.7 & 8.8 & 5.8 & 3.3 & 7.6 \\
		\cline{2-13}
		 & 16 ms & \multirow{2}{*}{2} & 128 & 11.6 & 12.8 & 13.7 & 14.3 & 17.0 & 14.2 & 12.0 & 9.3 & 13.1 \\
		 & 32 ms & & 256 & \textbf{12.2} & \textbf{13.3} & \textbf{14.0} & \textbf{14.7} & \textbf{17.4} & \textbf{14.5} & \textbf{12.4} & \textbf{9.9} & \textbf{13.5} \\
		\thline
	\end{tabular}
\end{table*}

\begin{table*}[!ht]
	\scriptsize
	\centering
	\caption{Comparison of different multi-channel pre-separation models on the simulated 6-mic circular array.}
	\label{tab:fixed-MC}
	\begin{tabular}{c|c|c|cccc|cccc|c|c}
		\thline
		\multirow{3}{*}{Model} & \multirow{3}{*}{\# of param.} & \multirow{3}{*}{\# of iter.} & \multicolumn{9}{c|}{SI-SDR (dB)} & \multirow{3}{*}{PESQ} \\
		\cline{4-12}
		& & & \multicolumn{4}{c|}{Speaker angle} & \multicolumn{4}{c|}{Overlap ratio} & \multirow{2}{*}{Average} & \\
		\cline{4-11}
		& & & $<$15\textdegree & 15-45\textdegree & 45-90\textdegree & $>$90\textdegree & $<$25\% & 25-50\% & 50-75\% & $>$75\% & & \\
		\hline
		Mixture & -- & -- & -0.5 & -0.4 & -0.4 & -0.4 & -0.4 & -0.4 & -0.5 & -0.4 & -0.4 & 1.35 \\
		\hline
		MC-TasNet-S & 1.3M & \multirow{2}{*}{--} & 7.6 & 7.8 & 8.3 & 8.4 & 13.0 & 9.0 & 6.3 & 3.7 & 8.0 & 1.54 \\
		MC-TasNet-L & 2.6M & & 8.2 & 8.5 & 8.8 & 9.1 & 13.4 & 9.7 & 7.0 & 4.4 & 8.6 & 1.59 \\
		\hline
		FaSNet-TAC-S & 2.1M & \multirow{2}{*}{--} & 7.6 & 9.8 & 11.4 & 12.2 & 14.1 & 11.2 & 9.0 & 6.6 & 10.2 & 1.77 \\
		FaSNet-TAC-L & 3.5M & & 8.3 & 10.4 & 11.8 & 12.6 & 14.6 & 11.7 & 9.4 & 7.3 & 10.8 & 1.81 \\
		\hline
		iFaSNet-S & 2.0M & \multirow{2}{*}{--} & 7.8 & 8.9 & 9.8 & 9.7 & 13.7 & 10.1 & 7.5 & 4.8 & 9.0 & 1.62 \\
		iFaSNet-L & 3.3M & & 8.2 & 9.7 & 10.5 & 10.4 & 14.2 & 10.6 & 8.2 & 5.7 & 9.7 & 1.67 \\
		\hline
		TD-GWF-FaSNet-TAC & \multirow{2}{*}{4.0M} & 1 & \textbf{11.2} & \textbf{13.2} & \textbf{14.3} & \textbf{15.1} & \textbf{17.3} & \textbf{14.5} & \textbf{12.2} & \textbf{9.7} & \textbf{13.4} & \textbf{2.03} \\
		(32 ms, 256 group) & & 2 & \textbf{11.9} & \textbf{14.0} & \textbf{15.2} & \textbf{15.9} & \textbf{17.8} & \textbf{15.2} & \textbf{13.2} & \textbf{10.6} & \textbf{14.2} & \textbf{2.17} \\
		\thline
	\end{tabular}
\end{table*}

\begin{table*}[!ht]
	\scriptsize
	\centering
	\caption{Comparison of different models on the performance of joint separation and dereverberation on the simulated 6-mic circular array.}
	\label{tab:fixed-MC-joint}
	\begin{tabular}{c|c|c|cccc|cccc|c|c}
		\thline
		\multirow{3}{*}{Model} & \multirow{3}{*}{\# of param.} & \multirow{3}{*}{\# of iter.} & \multicolumn{8}{c|}{SI-SDR (dB)} & &  \multirow{3}{*}{PESQ} \\
		\cline{4-12}
		& & & \multicolumn{4}{c|}{Speaker angle} & \multicolumn{4}{c|}{Overlap ratio} & \multirow{2}{*}{Average} & \\
		\cline{4-11}
		& & & $<$15\textdegree & 15-45\textdegree & 45-90\textdegree & $>$90\textdegree & $<$25\% & 25-50\% & 50-75\% & $>$75\% & & \\
		\hline
		Mixture & -- & -- & -0.9 & -0.9 & -0.8 & -0.7 & -0.9 & -0.8 & -0.8 & -0.8 & -0.8 & 1.27 \\
		\hline
		DPRNN-TasNet-S & 1.3M & \multirow{2}{*}{--} & 6.8 & 7.1 & 7.5 & 7.8 & 10.9 & 8.4 & 6.2 & 3.8 & 7.3 & 1.45 \\
		DPRNN-TasNet-L & 2.6M & & 7.2 & 7.6 & 8.0 & 8.2 & 11.2 & 8.8 & 6.7 & 4.3 & 7.7 & 1.50 \\
		\hline
		FaSNet-TAC-S & 2.1M & \multirow{2}{*}{--} & 6.8 & 8.6 & 10.1 & 10.6 & 11.9 & 10.0 & 8.2 & 6.0 & 9.0 & 1.61 \\
		FaSNet-TAC-L & 3.5M & & 7.2 & 9.1 & 10.5 & 11.2 & 12.3 & 10.4 & 8.6 & 6.7 & 9.5 & 1.68 \\
		\hline
		FD-MCWF-TasNet & \multirow{2}{*}{2.6M} & 1 & 8.8 & 9.0 & 9.5 & 9.7 & 12.5 & 10.4 & 8.3 & 5.9 & 9.3 & 1.65 \\
		(512 ms) & & 2 & 9.5 & 9.8 & 10.2 & 10.4 & 13.0 & 11.1 & 9.2 & 6.5 & 10.0 & 1.72 \\
		\hline
		TD-GWF-TasNet & \multirow{2}{*}{3.2M} & 1 & \textbf{10.0} & 10.8 & 11.7 & 12.1 & 13.6 & 12.1 & 10.5 & 8.3 & 11.1 & 1.76 \\
		(32 ms, 256 group) & & 2 & \textbf{10.6} & 11.6 & 12.3 & 13.1 & 14.2 & 12.8 & 11.4 & 9.1 & 11.9 & 1.87 \\
		\hline
	    TD-GWF-FaSNet-TAC & \multirow{2}{*}{4.0M} & 1 & 9.8 & \textbf{11.8} & \textbf{13.0} & \textbf{13.5} & \textbf{14.4} & \textbf{12.9} & \textbf{11.5} & \textbf{9.3} & \textbf{12.0} & \textbf{1.90} \\
		(32 ms, 256 group) & & 2 & \textbf{10.6} & \textbf{12.3} & \textbf{13.3} & \textbf{14.0} & \textbf{14.5} & \textbf{13.5} & \textbf{12.1} & \textbf{10.0} & \textbf{12.5} & \textbf{1.97} \\
		\thline
	\end{tabular}
\end{table*}

\begin{table*}[!ht]
	\scriptsize
	\centering
	\caption{Comparison of different models on the simulated ad-hoc array.}
	\label{tab:adhoc}
	\begin{tabular}{c|c|c|cccc|c|c}
		\thline
		\multirow{3}{*}{Model} & \multirow{3}{*}{\# of iter.}& \multirow{3}{*}{\# of mics} & \multicolumn{5}{c|}{SI-SDR (dB)} & \multirow{3}{*}{PESQ} \\
		\cline{4-8}
		& & & \multicolumn{4}{c|}{Overlap ratio} & \multirow{2}{*}{Average} \\
		\cline{4-7}
		& & & $<$25\% & 25-50\% & 50-75\% & $>$75\% &  \\
		\hline
		Mixture & -- & \multirow{15}{*}{2 / 4 / 6} & -0.4 / -0.5 / -0.5 & -0.5 / -0.4 / -0.4  & -0.4 / -0.5 / -0.4 & -0.4 / -0.6 / -0.5 & -0.4 / -0.5 / -0.4 & 1.34 / 1.36 / 1.35 \\
		\cline{1-2}\cline{4-9}
		DPRNN-TasNet-S & \multirow{2}{*}{--} & & 14.2 / 13.3 / 13.9 & 9.4 / 9.3 / 9.6 & 7.0 / 6.5 / 6.8 & 3.8 / 3.7 / 3.8 & 8.4 / 8.3 / 8.8 & 1.56 / 1.57 / 1.57  \\
		DPRNN-TasNet-L & & & 14.3 / 13.9 / 14.0 & 9.7 / 9.4 / 9.6 & 7.3 / 6.3 / 7.0 & 4.0 / 4.1 / 4.2 & 8.7 / 8.6 / 9.0 & 1.58 / 1.60 / 1.60 \\
		\cline{1-2}\cline{4-9}
		FaSNet-TAC-S & \multirow{2}{*}{--} & & 14.3 / 14.1 / 13.9 & 9.3 / 10.1 / 10.6 & 7.4 / 7.7 / 8.1 & 4.1 / 4.6 / 5.1 & 8.6 / 9.3 / 9.6 & 1.59 / 1.61 / 1.62 \\
		FaSNet-TAC-L & & & 14.9 / 15.0 / 14.4 & 10.3 / 10.9 / 11.4 & 8.1 / 8.7 / 9.0 & 4.8 / 6.4 / 6.4 & 9.4 / 10.4 / 10.5 & 1.66 / 1.69 / 1.70 \\
		\cline{1-2}\cline{4-9}
		iFaSNet-S & \multirow{2}{*}{--} &  & 14.9 / 15.7 / 15.6 & 10.4 / 11.5 / 12.0 & 8.3 / 9.0 / 9.5 & 4.8 / 6.6 / 7.4 & 9.4 / 10.8 / 11.4 & 1.65 / 1.78 / 1.78 \\
		iFaSNet-L & & & 15.2 / 15.9 / 16.0 & 10.7 / 11.8 / 12.5 & 8.5 / 9.4 / 9.9 & 5.2 / 7.3 / 8.0 & 9.7 / 11.2 / 11.8 & 1.68 / 1.81 / 1.83 \\
		\cline{1-2}\cline{4-9}
		FD-MCWF-TasNet & 1 & & 16.0 / 15.8 / 15.7 & 11.4 / 11.7 / 11.8 & 9.1 / 8.8 / 9.0 & 5.6 / 5.4 / 6.0 & 10.3 / 10.6 / 10.9 & 1.77 / 1.82 / 1.80 \\
		(512 ms) & 2 & & 16.6 / 16.8 / 16.5 & 12.4 / 12.7 / 12.8 & 10.0 / 10.2 / 10.3 & 6.8 / 7.0 / 7.0 & 11.3 / 11.8 / 11.9 & 1.86 / 1.94 / 1.90 \\
		\cline{1-2}\cline{4-9}
		TD-GWF-TasNet & 1 & & 15.4 / 16.2 / 16.4 & 11.6 / 12.3 / 13.0 & 9.3 / 9.8 / 10.3 & 6.8 / 7.5 / 7.7 & 10.6 / 11.5 / 12.1 & 1.74 / 1.83 / 1.83 \\
		(32 ms, 256 group) & 2 & & 16.5 / 17.1 / 17.4 & 12.0 / 13.3 / 14.2 & 9.8 / 10.2 / 11.8 & 6.7 / 8.3 / 8.7 & 11.1 / 12.3 / 13.3 & 1.77 / 1.91 / 1.94 \\
		\cline{1-2}\cline{4-9}
		TD-GWF-FaSNet-TAC & 1 & & \textbf{16.5} / 17.2 / 17.4 & 12.0 / 13.2 / 14.5 & 9.4 / 10.9 / 11.7 & 6.2 / 8.0 / 8.9 & 10.8 / 12.5 / 13.3 & 1.77 / 1.92 / 1.94 \\
		(32 ms, 256 group) & 2 & & 16.7 / 17.5 / 17.8 & 12.1 / 13.7 / 14.9 & 9.9 / 11.5 / 12.2 & 6.9 / 9.1 / 9.8 & 11.2 / 13.1 / 13.9 & 1.80 / 1.96 / 1.99 \\
		\cline{1-2}\cline{4-9}
		TD-GWF-iFaSNet & 1 & & 16.0 / \textbf{17.6} / \textbf{17.6} & \textbf{12.2} / \textbf{13.9} / \textbf{14.8} & \textbf{9.9} / \textbf{11.3} / \textbf{12.2} & \textbf{6.9} / \textbf{9.3} / \textbf{10.0} & \textbf{11.1} / \textbf{13.1} / \textbf{13.9} & \textbf{1.80} / \textbf{1.99} / \textbf{2.00} \\
		(32 ms, 256 group) & 2 & & \textbf{17.1} / \textbf{17.9} / \textbf{18.3} & \textbf{12.8} / \textbf{14.4} / \textbf{15.3} & \textbf{10.4} / \textbf{12.0} / \textbf{13.0} & \textbf{7.8} / \textbf{10.0} / \textbf{11.0} & \textbf{11.9} / \textbf{13.7} / \textbf{14.6} & \textbf{1.90} / \textbf{2.09} / \textbf{2.13} \\
		\thline
	\end{tabular}
\end{table*}

We then use the LUT encoder/decoder as the default signal transform and further investigate the effect of larger window sizes and number of groups. Table~\ref{tab:fixed-config} provides the performance comparison of LUT-based systems with up to 512~ms window size (i.e., up to $P=8192$) and up to $V=P/2$ groups. We first notice that the LUT encoder/decoder still outperforms identity encoder/decoder with 8 and 16~ms windows, and the performance of identity encoder/decoder system with 16~ms window is not improved compared to the one with 8~ms window. This shows that unlike conventional frequency-domain beamformers, increasing the window size in TD-GWF with identity encoder/decoder does not always lead to a performance improvement. We then find that a relatively better configuration is found at 32~ms window ($P=512$) and $V=256$ groups, which significantly outperforms the results in Table~\ref{tab:fixed-enc} with a 4~ms window size. Given that the TD-GWF module with $N=P=512$ and $V=256$ only contains $(MN/V\times N/V)\times V=(12 \times 2)\times 256=6144$ filter coefficients and the calculation of filter coefficients in the 256 groups can be done in parallel, the result shows that TD-GWF is able to achieve significantly better performance than FD-MCWF without significantly increasing the computational complexity.

\subsection{Performance of Multi-channel Separation Systems for the Pre-separation Module}

All systems above use a single-channel pre-separation module. Here we conduct experiments to see if a multi-channel pre-separation module can further improve the overall performance. Table~\ref{tab:fixed-MC} provides the performance of three multi-channel benchmark systems and one TD-GWF-based sequential beamforming pipeline on the fixed geometry array. Note that all selected multi-channel systems in Section~\ref{sec:model-config} only perform separation on the reference channel, while they utilize cross-channel information in different ways. We find that inserting the TD-GWF module to the best-performed multi-channel system, which is the FaSNet-TAC system in our comparison, can still significantly improves the overall performance. Moreover, the performance of the best reported system here is better than the one using the single-channel DPRNN-TasNet system for pre-separation module in Table~\ref{tab:fixed-enc}, which confirms that improving the performance of the pre-separation system can lead to a better overall performance.

\subsection{Performance of Different System Configurations on the Joint Separation and Dereverberation Task}

The experiments above all use the reverberant SOIs as the training target. Here we modify the training target to the direct-path SOIs to see if the systems can also benefit from TD-GWF when jointly performing separation and dereverberation. In our case, the direct-path RIR is defined as ±6~ms of the first peak in the RIR, and the direct-path SOI is obtained by convolving the direct-path RIR with the original source. Table~\ref{tab:fixed-MC-joint} shows the performance of single-channel and multi-channel benchmark systems as well as their TD-GWF-based sequential beamforming pipelines on the fixed geometry array. We can see that compared to the benchmark systems, adding the TD-GWF modules achieve a similar performance improvement to the separation-only task, which confirms the effectiveness of TD-GWF in both separation and dereverberation.

\subsection{Performance of Different System Configurations on Ad-hoc Array}

Finally we evaluate the effect of TD-GWF on the ad-hoc array. Table~\ref{tab:adhoc} provides the performance of the benchmark systems as well as their sequential beamforming pipelines, and the results are reported for different numbers of microphones. Compared to the results on the fixed geometry array, we again observe that TD-GWF is able to significantly improve the separation performance with various single-channel and multi-channel pre-separation modules, and TD-GWF is also able to achieve higher performance improvement than FD-MCWF. The results prove that TD-GWF has the potential to replace conventional frequency-domain beamformers in a wide range of microphone array scenarios and task configurations.

\section{Conclusion and future works}
\label{sec:conclusion}
In this paper, we proposed time-domain real-valued generalized Wiener filter (TD-GWF), a simple yet effective replacement to the conventional frequency-domain beamformers in the sequential neural beamforming pipelines. Unlike conventional time-domain beamformers, TD-GWF applied a learnable 2-D representation to the 1-D waveform to generate a 2-D representation and split the representation into non-overlapped groups. The filter coefficients were estimated in different groups in parallel and applied to the noisy observation to generate the estimated 2-D representation of the target source. Such group-splitting operation was able to not only reduce the computational complexity but also improve the separation performance. Experiment results showed that TD-GWF not only achieved better oracle performance than conventional frequency-domain beamformers on signal quality measurements, but also performed consistently better in the sequential neural beamforming pipelines when replacing the frequency-domain beamformers on various microphone array scenarios and task configurations.

There are multiple things we leave as future works. First, unlike conventional filter-and-sum beamformers where a beampattern can be calculated and visualized, TD-GWF does not have a clear definition of ``beampattern'' as the learnable signal transform is not orthonormal and the number of groups does not equal to the window size. A better way to understand how TD-GWF performs spatial filtering is thus necessary. Second, the proposed definition of TD-GWF requires the entire utterance to be available, and how to modify it to support streaming calculation is important for real-world applications. Third, as conventional frequency-domain beamformers are widely used in ASR tasks, it is important to evaluate the performance of TD-GWF in different multi-channel ASR systems. Fourth, since GWF can be applied to any multi-channel signals such as biological or multi-antenna signals, it is thus interesting to investigate its potential in other types of data.

\bibliographystyle{IEEEbib}
\bibliography{refs}

\end{document}